\documentclass[11pt]{article}
\usepackage{epsf,amsmath,amssymb,axodraw,graphicx,scalefnt,subfigure,ulem}
\usepackage{cite}
\usepackage{color}
\usepackage{rotating}

\newcommand{\pdf}{{\abbrev PDF}}
\newcommand{\qcd}{{\abbrev QCD}}

\newcommand{\abbrev}{\scalefont{.9}}

\newcommand{\muF}{\mu_{\rm F}}
\newcommand{\muR}{\mu_{\rm R}}
\newcommand{\mhiggs}{M_{\rm H}}

\newcommand{\eqn}[1]{Eq.\,(\ref{#1})}
\newcommand{\fig}[1]{Fig.\,\ref{#1}}

\newcommand{\dd}{{\rm d}}

\newcommand{\lhc}{{\abbrev LHC}}
\newcommand{\sm}{{\abbrev SM}}
\newcommand{\mssm}{{\abbrev MSSM}}
\newcommand{\susy}{{\abbrev SUSY}}

\newcommand{\lo}{{\abbrev LO}}
\newcommand{\nlo}{{\abbrev NLO}}
\newcommand{\nnlo}{{\abbrev NNLO}}

\newcommand{\msbar}{\overline{\mbox{\abbrev MS}}}

\newcommand{\bld}[1]{\boldmath{$#1$}}

\date{}
\title{\vspace*{-6em}
  \begin{flushright}
    {\sf\small \phantom{{\bf 29/07/2010} ---} July 2010 --- WUB/10-20}
  \end{flushright}
  \vspace*{2em} Higgs plus jet production in bottom quark annihilation
  at next-to-leading order} \author{Robert V. Harlander, Kemal
  J. Ozeren, Marius Wiesemann\\[2em] {\it Fachbereich C, Bergische
    Universit\"at Wuppertal}\\{\it 42097 Wuppertal,
    Germany}\\
  {\small\tt harlander@physik.uni-wuppertal.de}\\[-.3em]
  {\small\tt ozeren@physik.uni-wuppertal.de}\\[-.3em]
  {\small\tt m.wiesemann@uni-wuppertal.de}
}
\begin{document}
\maketitle

\begin{abstract}
The cross section for Higgs+jet production in bottom quark annihilation
is calculated through \nlo{} \qcd{}. The five-flavour scheme is used to
derive this contribution to the Higgs+jet production cross section which
becomes numerically important in the \mssm{} for large values of
$\tan\beta$. We present numerical results for a proton collider with
14\,TeV center-of-mass energy. The \nlo{} matrix elements for
$\dd\sigma/\dd p_T$ are then combined with the total inclusive cross
section in order to derive the integrated cross section with a maximum
cut on $p_T$ at next-to-next-to-leading order.
\end{abstract}

\section{Introduction}

The Higgs mechanism~\cite{Guralnik:1964eu,Englert:1964et,Higgs:1964pj}
plays a central role in both the Standard Model
(\sm{})~\cite{Glashow:1961tr,Weinberg:1967tq,Salam:1968??} and its
supersymmetric extensions~\cite{Nilles:1983ge}. The gauge bosons and
quarks acquire masses through interactions with Higgs fields.  Up to
now, the search for the physical Higgs boson has been unsuccessful but
has led to the exclusion of a certain Higgs mass
range~\cite{Barate:2003sz,Aaltonen:2010yv}. In combination with the fits
of electro-weak precision data to higher order perturbative calculations
this leads to a rather small range of allowed values for a \sm{} Higgs
boson mass~\cite{lepewwg}.

Supersymmetric (\susy{}) theories require an enlarged Higgs sector. The
minimal \susy{} extension of the \sm{} leads to five physical Higgs
bosons. Due to the larger number of free parameters in \susy{}, the
sensitivity of experimental data to Higgs bosons is weaker than in the
\sm{}~\cite{Schael:2006cr,Benjamin:2010xb}.

The Large Hadron Collider (\lhc{}) is expected to find a Higgs boson if
it exists. To do this, various production and decay channels must be
considered. The relative utility of each channel depends strongly on the
Higgs mass and couplings. While in the \sm{} gluon fusion is the Higgs
production process with the largest cross section by far, in \susy{}
theories with large $\tan\beta$, Higgs production in association with
bottom quarks is dominant~(for reviews, see
Refs.\,\cite{Djouadi:2005gi,Djouadi:2005gj}; detailed studies of the
relative importance of gluon fusion and bottom annihilation have been
performed in
Ref.\,\cite{Belyaev:2005ct,Belyaev:2005nu,Brein:2010xj}). This is
because in this region of the \susy{} parameter space the
$Hb\overline{b}$ coupling is enhanced relative to the \sm{}.

\begin{figure}
\centering
\subfigure[]{\label{fig::four}\includegraphics[width=6em]{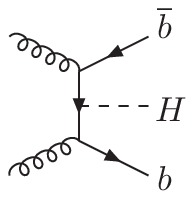}}
\hspace{1.5in}
\subfigure[]{\label{fig::five}\includegraphics[width=7.5em]{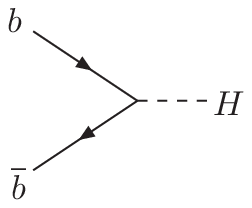}}
\caption{Leading order diagrams for associated $b\overline{b}h$
  production in the (a) four and (b) five flavour scheme.}
\end{figure}

Assuming that only four quark flavours and the gluon make up the proton
(the so-called ``four-flavour scheme'' or {\abbrev 4FS}), the dominant
leading order Feynman diagram for this process is shown in
\fig{fig::four}. If the bottom quark was massless, integration over
phase space would lead to divergences arising from the kinematical
region where one or both bottom quarks are collinear to the incoming
partons. The bottom quark mass regulates these divergences, but they
still leave traces in terms of logarithms of the form
$\ln(m^2_b/m^2_H)$.  Such logarithms lead to large perturbative
coefficients, so ideally one would like to resum them.  This can be
achieved by considering this process in the five-flavour scheme
({\abbrev 5FS})~\cite{Barnett:1987jw,Dicus:1989cx}, i.e.\ by introducing
bottom quark \pdf{}s (parton density functions).\footnote{In fact, this
  has been the default for all modern \pdf{} sets; see
  Ref.\,\cite{Martin:2010db} though.} Now that the $b$ quarks can appear
in the initial state, the leading order process is changed to that in
\fig{fig::five}. We note that the scheme choice amounts merely to a
re-ordering of the perturbative series. Of course, when truncated at a
finite order, results obtained in either scheme will differ, with the
difference being formally of higher order in $\alpha_s$.

However, it was found that the difference between the inclusive cross
section in the {\abbrev 4FS} and the {\abbrev 5FS} differs by roughly a
factor of five when evaluated at $\muF=\muR=\mhiggs$, where $\muF/\muR$
is the factorization/renormalization scale.  This remains true also at
\nlo{} \qcd{} which was calculated for the {\abbrev 5FS} in
Ref.\,\cite{Dicus:1998hs,Maltoni:2003pn}, and for the {\abbrev 4FS} in
Ref.\,\cite{Dittmaier:2003ej,Dawson:2003kb}.  It was thus proposed in
Refs.\,\cite{Rainwater:2002hm,Plehn:2002vy,Maltoni:2003pn,Boos:2003yi}
that when using the five flavour scheme the appropriate scale choice is
$m_H/4$.

A weakness of the {\abbrev 5FS} is that it neglects the contribution
from large-$p_T$ bottom quarks at leading order.  This is taken into
account only at \nnlo{} and higher (note that the \lo{} set of Feynman
diagrams in the {\abbrev 4FS} is part of the \nnlo{} set in the {\abbrev
  5FS}). Indeed, the factorization scale dependence at \nnlo{} is very
flat~\cite{Harlander:2003ai} and seems to confirm the ``natural'' scale
choice at lower orders of $\muF=\mhiggs/4$.

It has been pointed out a long time ago that it can be advantageous to
consider the $H+$jet process instead of the fully inclusive production
when searching for the Higgs boson~\cite{Ellis:1987xu}. The $gg\to
H+$jet cross section is known at \lo{}, including the full top and
bottom quark mass dependence, both in the
\sm{}~\cite{Ellis:1987xu,Field:2003yy,Brein:2003df} and in the
\mssm{}~\cite{Brein:2003df}.  \nlo{} \qcd{} corrections are known in the
heavy-top
limit~\cite{deFlorian:1999zd,Ravindran:2002dc,Glosser:2002gm}. It is
expected that a very good approximation of the \mssm{} effects can be
obtained by simply replacing the corresponding Wilson coefficient of the
effective $ggh$ coupling with its \mssm{}
expression~\cite{Harlander:2003bb,Harlander:2004tp,Degrassi:2008zj}, at
least as long as $\tan\beta$ is not too large. Otherwise, bottom loop
effects which are not covered in the heavy-top limit will be important.
Resummation effects for small and large $p_T$ of the Higgs boson have
been treated in Refs.\,\cite{Kulesza:2003wn,Bozzi:2003jy,Bozzi:2005wk,%
  deFlorian:2005rr}.

As mentioned already above, for large $\tan\beta$, it is essential to
also take bottom quark annihilation into account. It is well known that
the corresponding \qcd{} corrections can be large. In this paper
  we present them for distributions of the Higgs boson. Since it has
been shown for the inclusive cross section that the dominant \susy{}
effects can be approximated to high accuracy by an effective $b\bar bh$
coupling~\cite{Dittmaier:2006cz}, our results are directly applicable to
the \mssm{} by a trivial overall rescaling.

In Ref.~\cite{Campbell:2002zm}, a related quantity, namely the Higgs
production cross section in association with a single tagged $b$ quark
was studied in the {\abbrev 5FS}. Tagging a $b$ may be useful for
measuring the $b\bar bh$ Yukawa coupling $y_b$, for example. In this
paper we consider Higgs production without necessarily tagging a final
state $b$. That is, we consider the $b\overline{b}$ initial state, and
its associated sub-channels, as a contribution to the inclusive
Higgs+jet cross section. We present the $p_T$ and $y$ distributions, and
study the scale dependence of the cross section and its component
channels. These results are valid to \nlo{} in \qcd{} perturbation
theory. They will be presented in Section~\ref{sec::nlo}.

Using the knowledge of the total inclusive cross section at
\nnlo{}~\cite{Harlander:2003ai}, we can then use our result for \nlo{}
$H+$jet production in order to derive the \nnlo{} cross section with
upper cuts on $p_T$. This will be described in Section~\ref{sec::nnlo}.

\section{Next-to-leading order cross section for \bld{\sigma(b\bar b\to h+}jet)}
\label{sec::nlo}

\begin{figure}
\centering
\subfigure[]{\label{fig::lead}\includegraphics[width=8em]{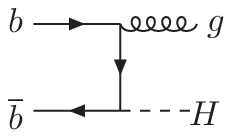}}
\hspace{1.5in}
\subfigure[]{\label{fig::leadgb}\includegraphics[width=8em]{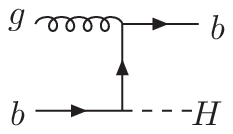}}
\caption{Representative diagrams for each of the two leading order channels.}
\label{fig::leaddias}
\end{figure}

The generic leading order diagrams to Higgs plus jet production are
shown in \fig{fig::leaddias}. At \nlo{} each of these receives virtual
corrections, and in addition we must include the real emission
contributions, which induce also other initial states. The full list of
processes at \nlo{} is: $b\overline{b}\to gH$ and $bg\to bH$ at one
loop, $b\overline{b} \to ggH$, $b\overline{b}\to b\overline{b}H$,
$b\overline{b}\to q\overline{q}H$, $gb \to gbH$, $bq\to bqH$,
$q\overline{q}\to b\overline{b}H$, $gg \to b\overline{b}H$, $bb\to bbH$
at tree level, where $q$ denotes any of the light quarks $u,d,s,c$. It
is understood that the charge conjugated processes must also be
included. Formally, the virtual contributions include diagrams where the
Higgs boson is radiated off a closed bottom or top quark loop. The
former lead to terms $\sim \alpha_s^2 y_b^2\cdot m_b^2/\mhiggs^2$,
however, which is neglected throughout our calculation, and in the
spirit of Refs.\,\cite{Campbell:2002zm,Harlander:2003ai}, we discard the
latter which are proportional to the top Yukawa coupling. They
are separately finite and gauge invariant and could simply be added to
our results, once the ratio of top and bottom Yukawa coupling is known.

The \nlo{} calculation of a process as the one considered here is by now
standard.  We apply the dipole subtraction method~\cite{Catani:1996vz}
in order to cancel the infra-red poles between the virtual and the real
radiation contributions in the $b\bar b$ and $bg$ processes. Introducing
the $\alpha$-parameter for restricting the dipole phase
space~\cite{Nagy:2003tz,Nagy:1998bb} not only improves the numerical
integration, but also serves as a welcome check through the requirement
of $\alpha$-independence of the final result. Furthermore, our result
for the virtual corrections agrees with the result of
Ref.\,\cite{Campbell:2002zm}.  The leading logarithmic behaviour
  at small $p_T$ can be checked numerically against the resummed
  expression of Ref.\,\cite{Belyaev:2005bs}.  The most important check,
however, is the numerical comparison to a fully analytic evaluation of
the $p_T$ distribution to be published elsewhere~\cite{Kemal}.

In order to avoid the infra-red divergence at low Higgs transverse
momenta $p_T$, we cut contributions from $p_T<30$\,GeV in this section.
For our numerical analysis we use the following set of input
parameters. The \pdf{}s are taken from the {\abbrev MSTW}2008
set~\cite{Martin:2009iq}, and the \qcd{} coupling is accordingly set to
$\alpha_s(M_Z) = 0.13939$ at \lo{}, and $\alpha_s(M_Z) = 0.12018$ at
\nlo{}. The $b\bar bh$ coupling, for which we assume the \sm{}
expression $m_b/v$ ($v=246.22$\,GeV), is evaluated with the running bottom
quark mass $m_b(\muR)$ defined in the $\msbar{}$ scheme, with an input
value $m_b(m_b) = 4.2$ GeV~\cite{Amsler:2008zzb}. Our default value for
the Higgs mass is $\mhiggs=120$\,GeV.

\begin{figure}
  \begin{center}
    \begin{tabular}{c}
      \subfigure[]{\includegraphics[width=.5\textwidth]{%
          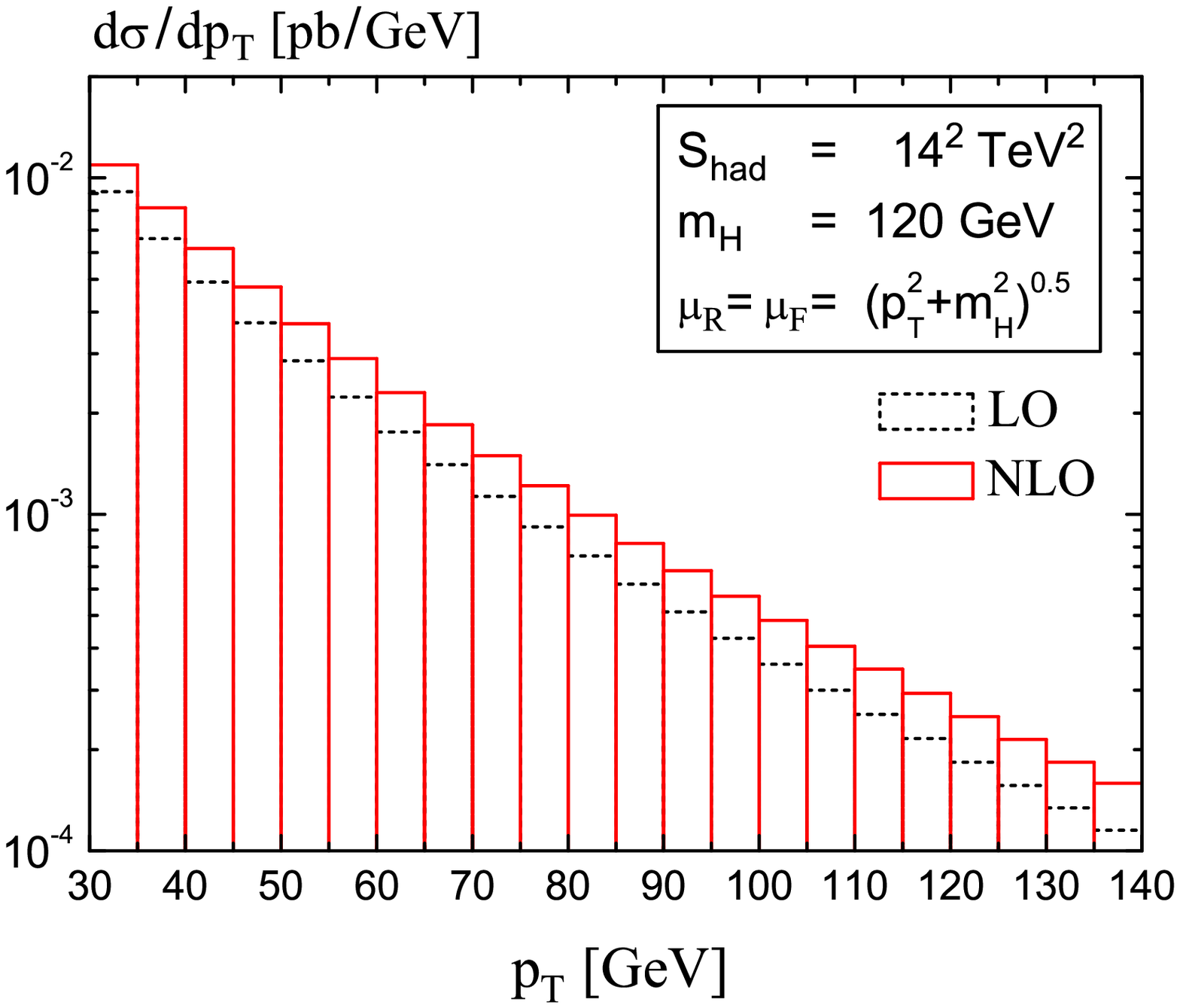}}
      \subfigure[]{\includegraphics[width=.5\textwidth]{%
          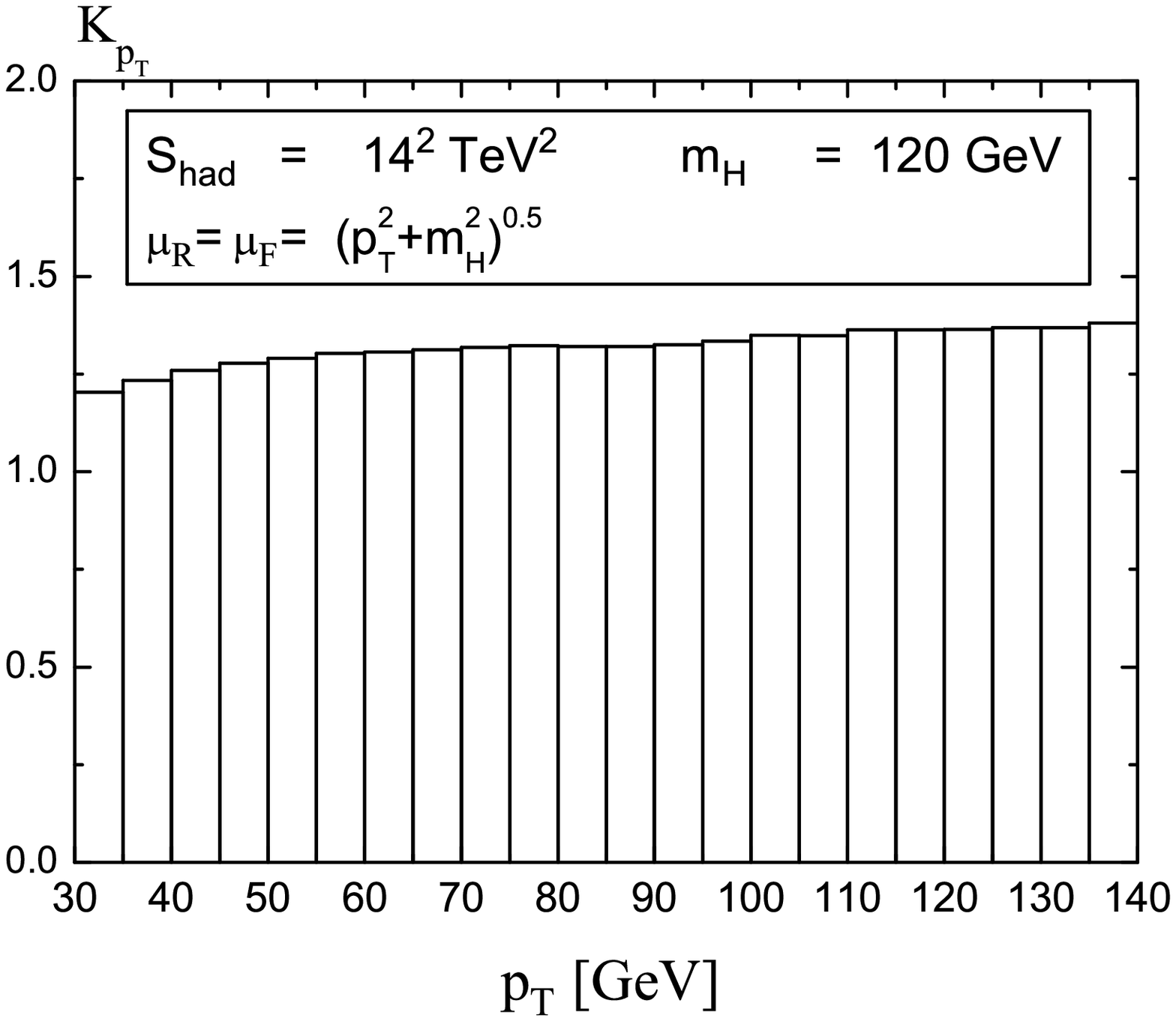}}
    \end{tabular}
    \caption[]{\label{fig::pt} (a) Higgs transverse momentum distribution at
      \lo{} (dashed) and \nlo{} (solid); (b) corresponding K-factor.}
  \end{center}
\end{figure}

\begin{figure}
  \begin{center}
    \begin{tabular}{c}
      \subfigure[]{\includegraphics[width=.5\textwidth]{%
          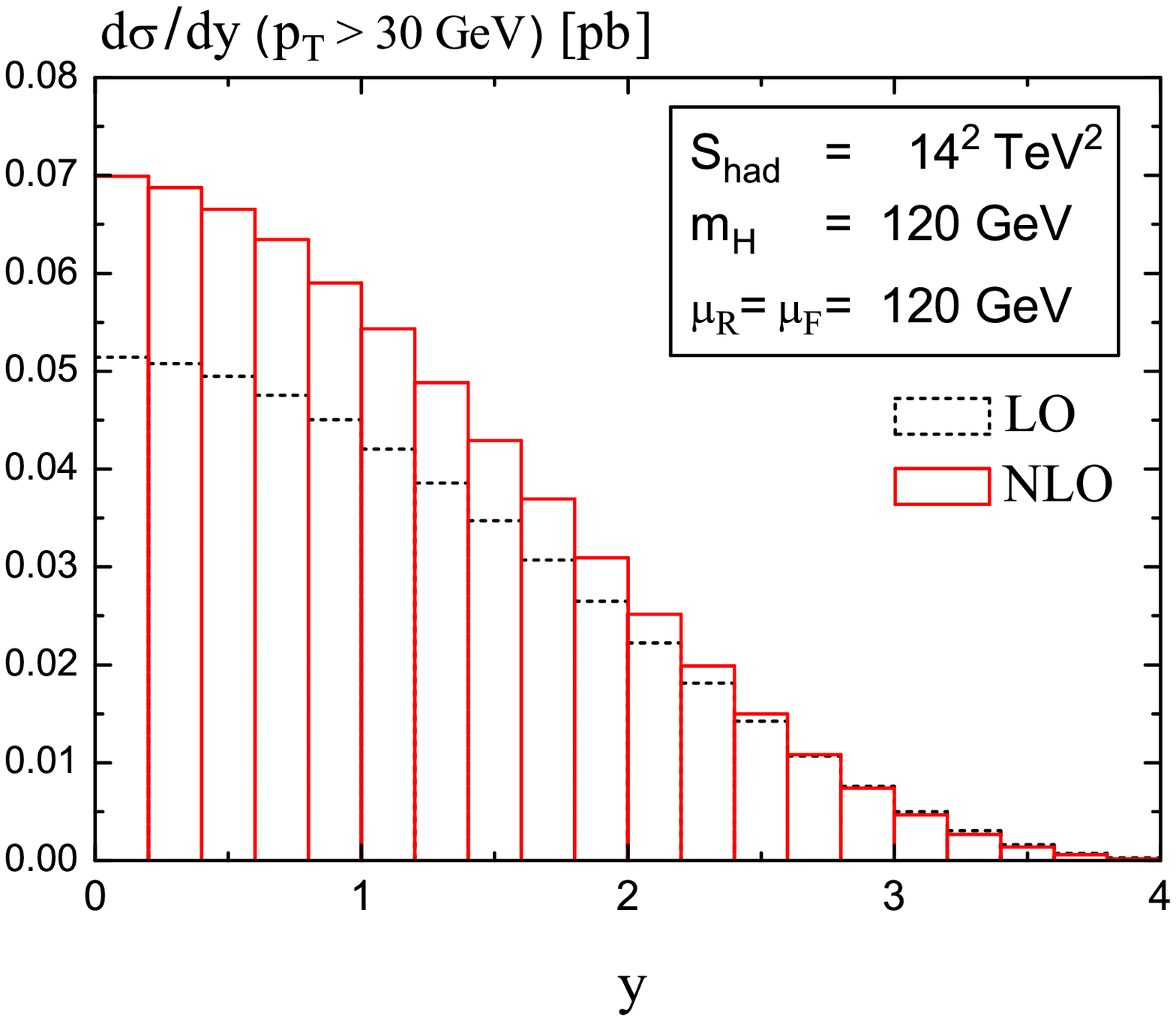}}
      \subfigure[]{\includegraphics[width=.5\textwidth]{%
          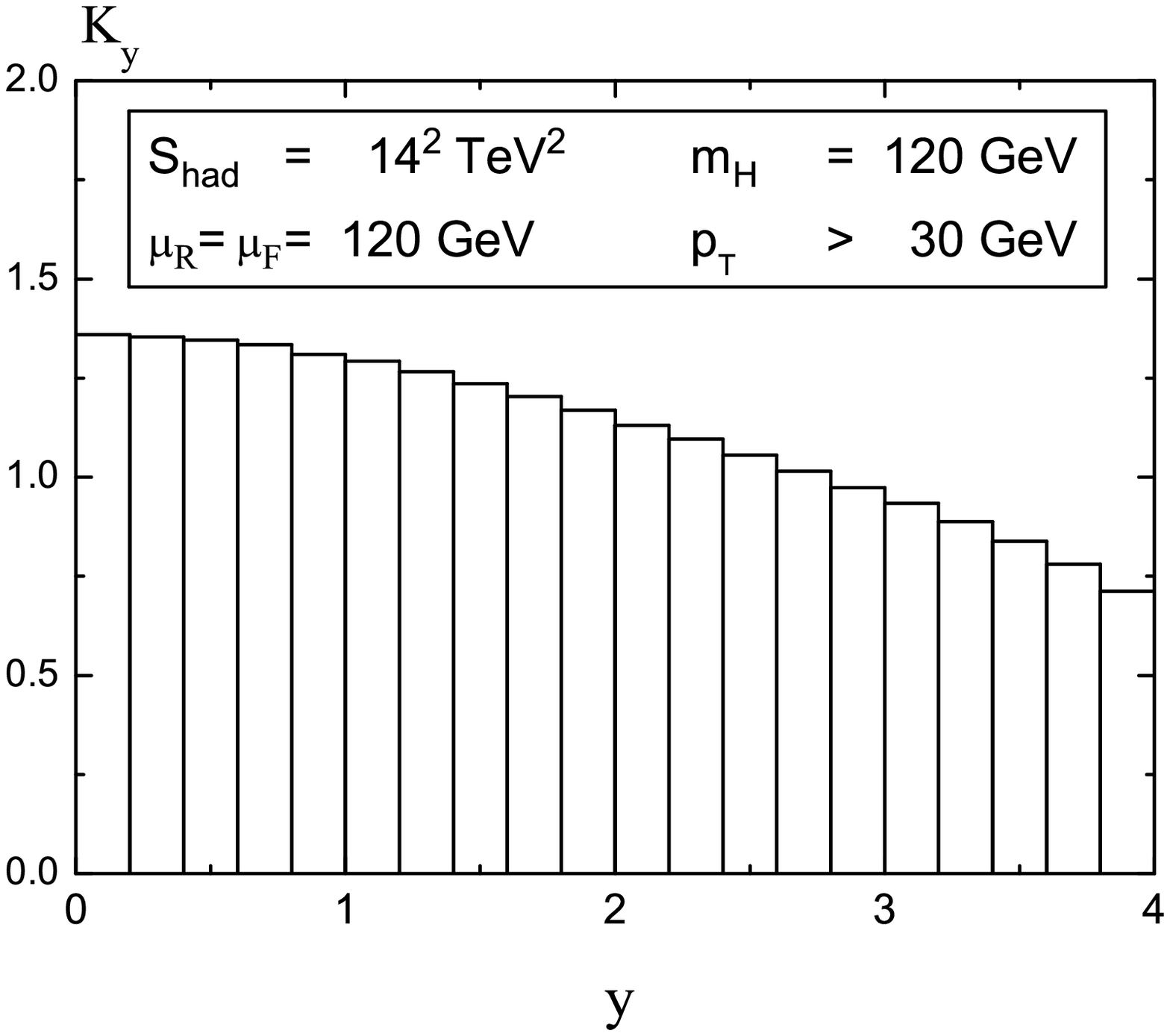}}
    \end{tabular}
    \caption[]{\label{fig::y} (a) Higgs rapidity distribution at \lo{}
      (dashed) and \nlo{} (solid); (b) corresponding K-factor. Since
      the distribution is symmetric around $y=0$, only positive values
      of $y$ are shown.}
  \end{center}
\end{figure}

In \fig{fig::pt} and \fig{fig::y} we show the transverse momentum and
rapidity distributions of the Higgs boson, both at \lo{} and \nlo{}, and
the corresponding $K$-factors
\begin{equation}
\begin{split}
K_{p_T} \equiv \frac{(\dd\sigma/\dd
  p_T)_{\rm\nlo}}{(\dd\sigma/\dd p_T)_{\rm\lo}}\,,\qquad
K_{y} \equiv \frac{(\dd\sigma/\dd
  y)_{\rm\nlo}}{(\dd\sigma/\dd y)_{\rm\lo}}\,,
\label{eq::kfacs}
\end{split}
\end{equation}
where
\begin{equation}
\begin{split}
y = \frac{1}{2}\ln\frac{E+p_z}{E-p_z}\,,
\end{split}
\end{equation}
and $E$ and $p_z$ are the energy and longitudinal component of the Higgs
boson in the lab frame. The choice of the renormalization and
factorization scales is given in the plots. We remark that the
numerator/denominator in \eqn{eq::kfacs} is evaluated with \nlo{}/\lo{}
\pdf{}s and couplings.  Both for the $p_T$ and the $y$ distribution, the
dependence of the $K$ factors on $p_T$ and $y$ is very similar to what
is observed for the gluon fusion
channel~\cite{deFlorian:1999zd,Ravindran:2002dc,Glosser:2002gm}:
$K_{p_T}$ is rather flat over the considered $p_T$ interval, while $K_y$
drops mildly towards larger values of the rapidity.  The absolute size
of the corrections is significantly smaller than in the gluon fusion
case though. Note that the $K$ factors for the distributions
  cannot be immediately deduced from the one for the inclusive cross
  section for $b\bar b\to H+X$ due to the strong $\muF$ dependence at
  \lo{}, cf.\,Ref.\,\cite{Harlander:2003ai} and \fig{fig::ptcut}\,(b).

\begin{figure}
  \begin{center}
    \begin{tabular}{c}
      \subfigure[]{\includegraphics[width=.5\textwidth]{%
          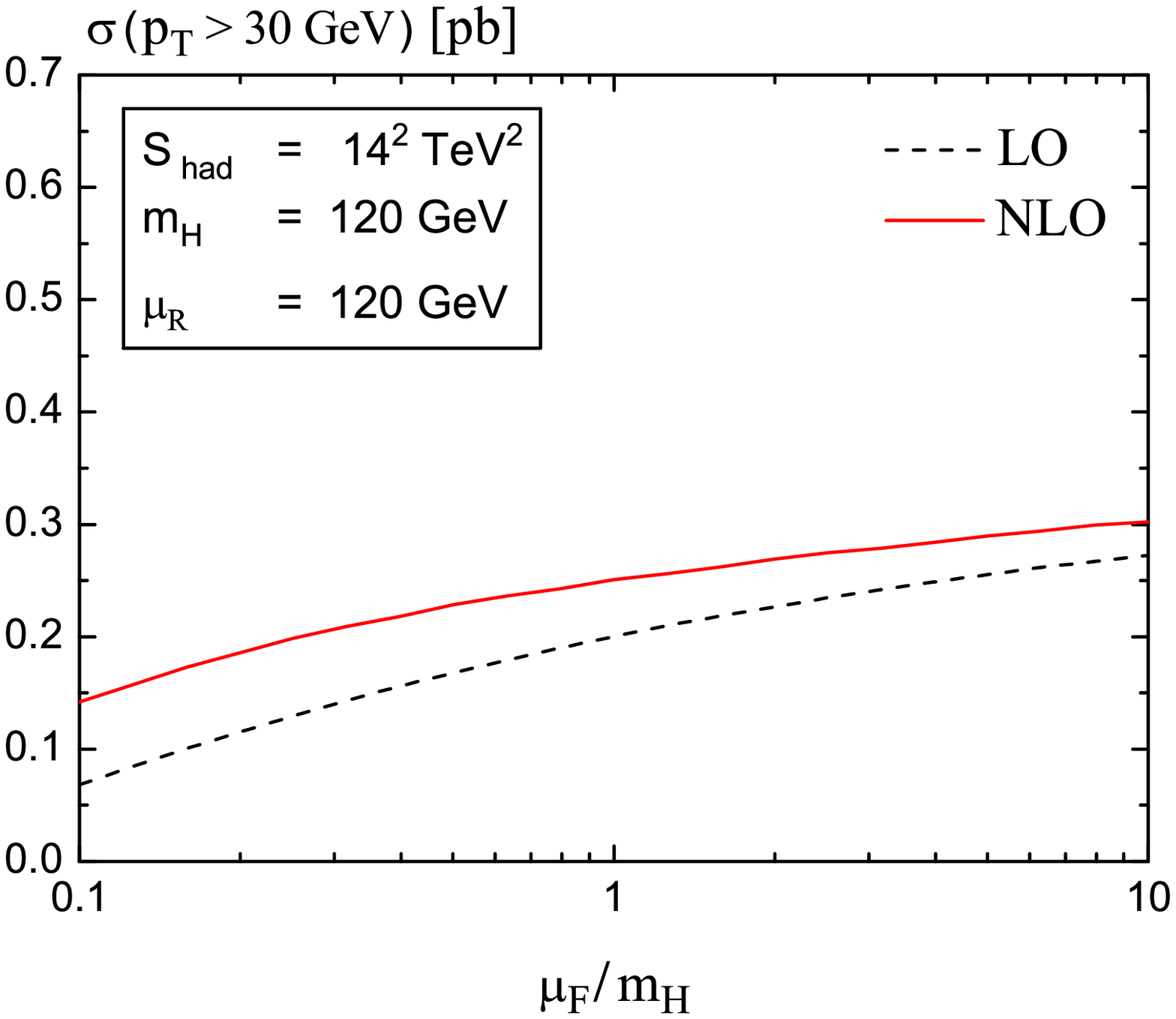}}
      \subfigure[]{\includegraphics[width=.5\textwidth]{%
          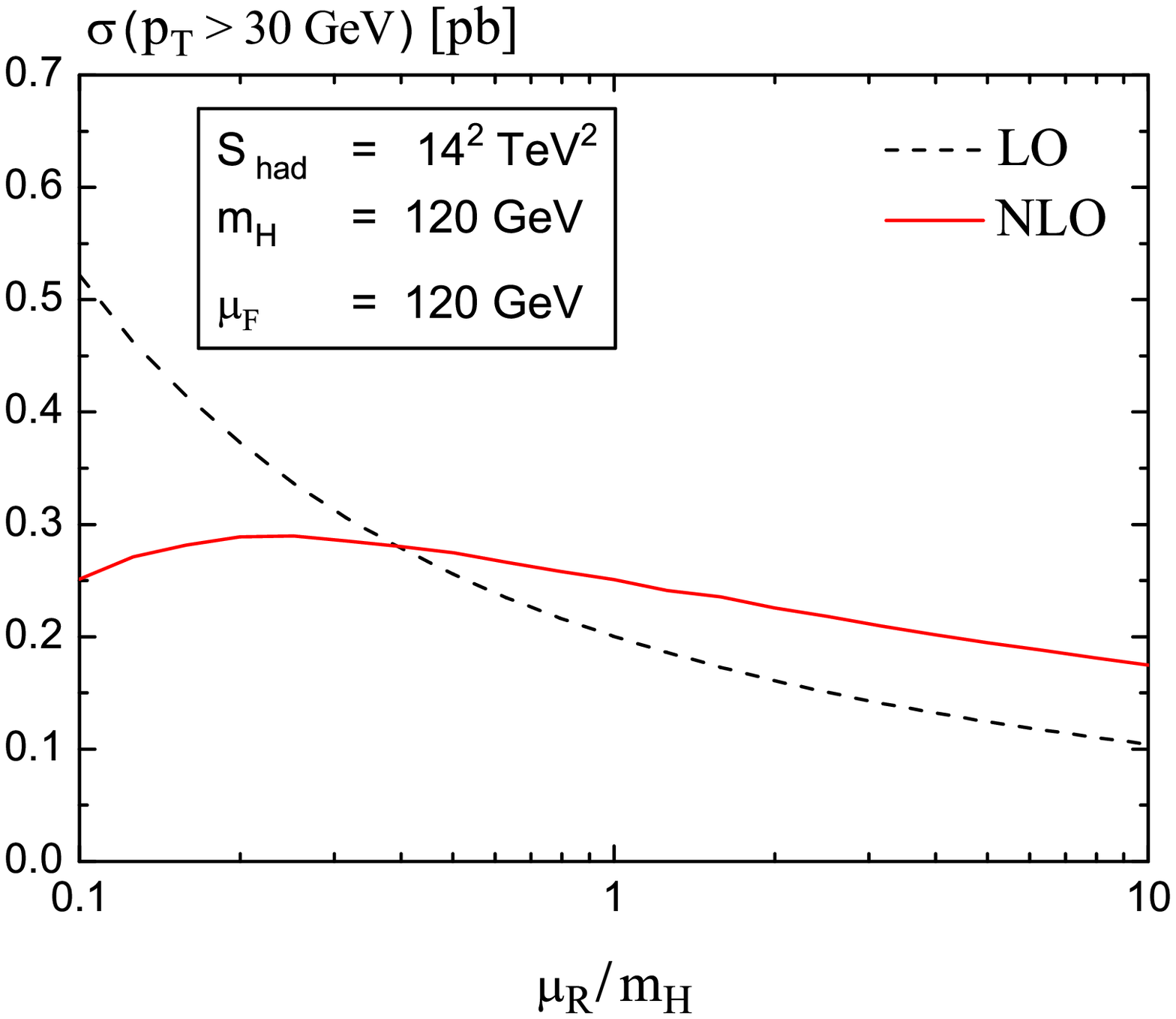}}\\
      \subfigure[]{\includegraphics[width=.5\textwidth]{%
          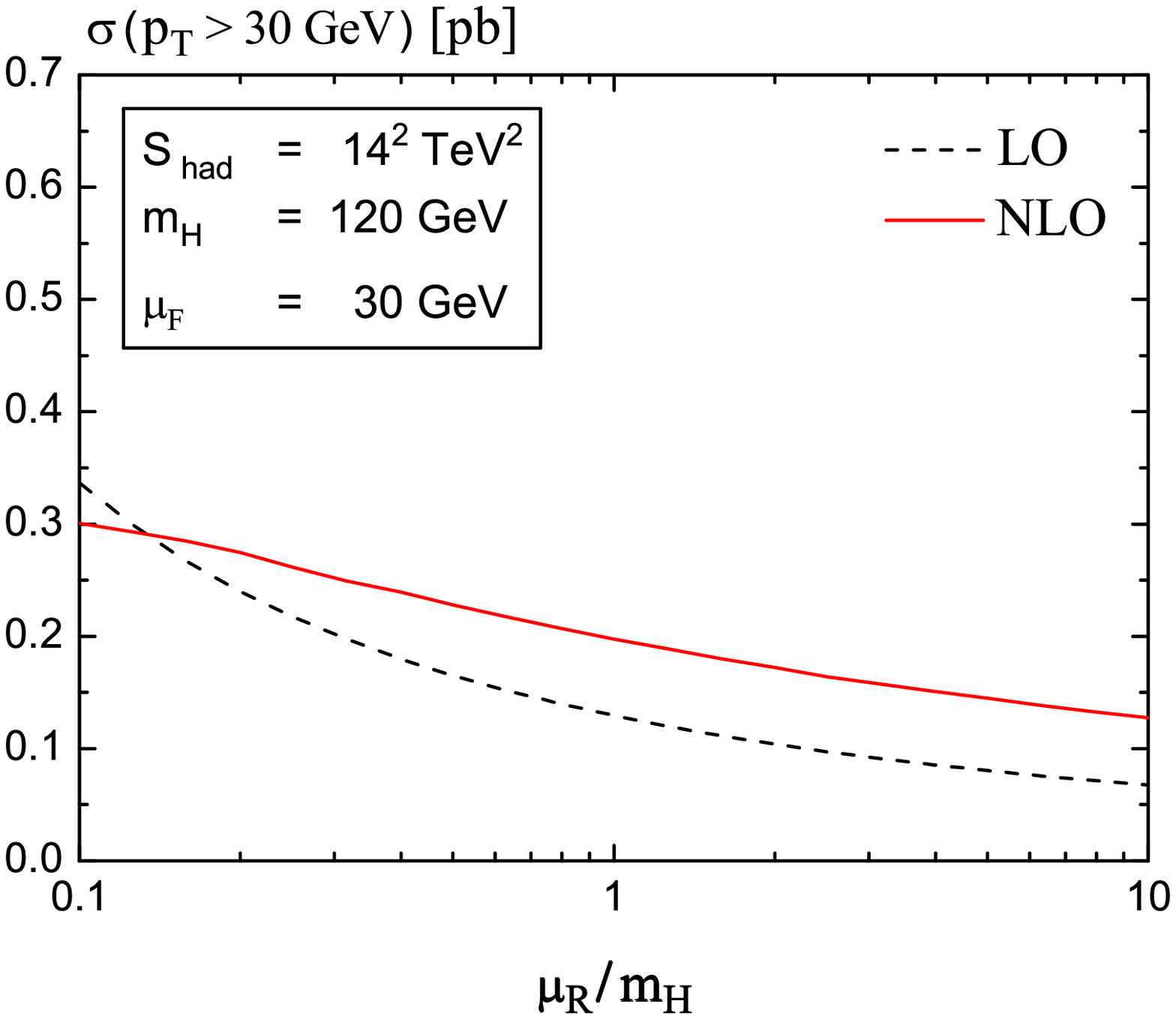}}
      \subfigure[]{\includegraphics[width=.5\textwidth]{%
          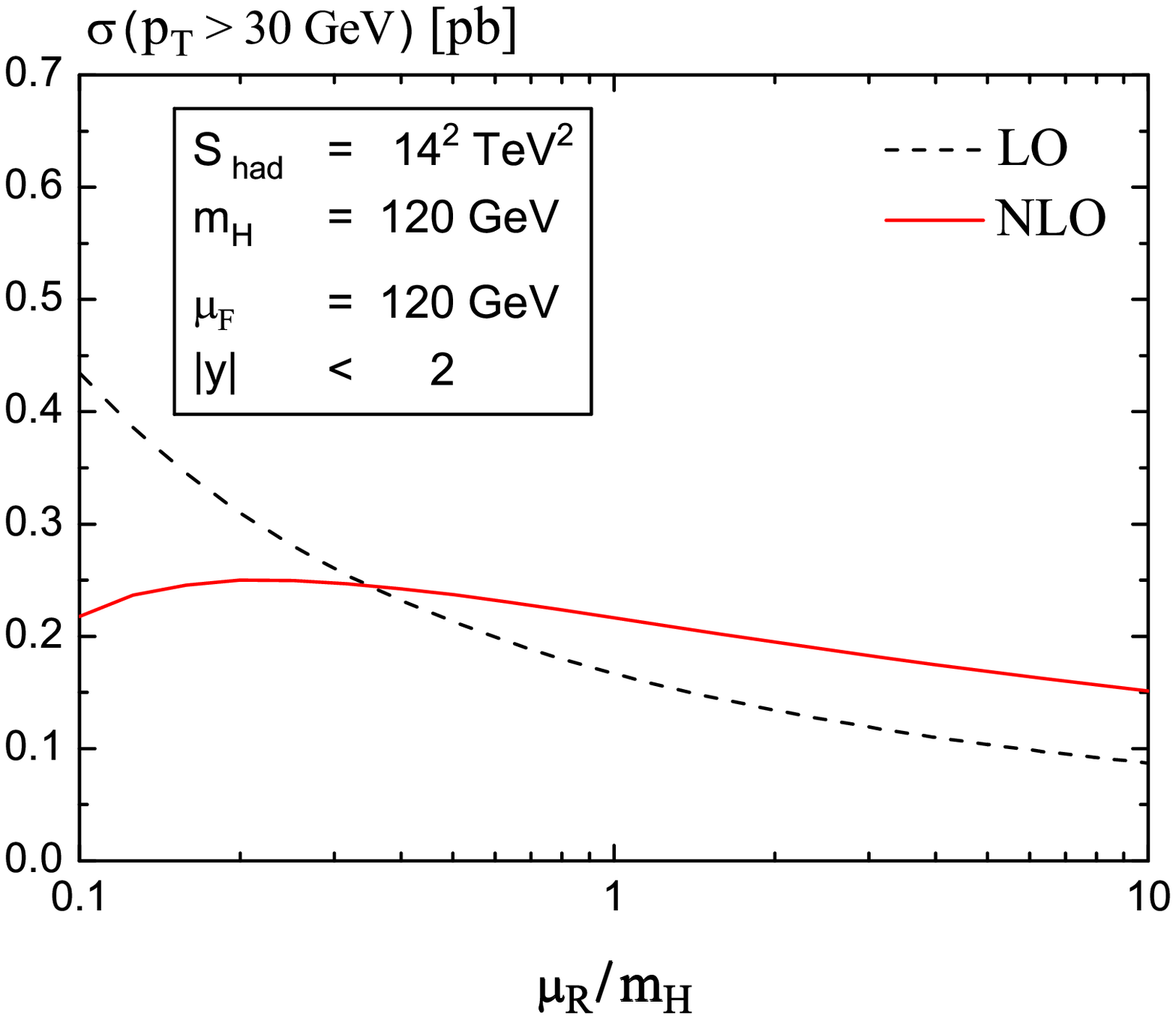}}
    \end{tabular}
    \caption[]{\label{fig::ptmin} Scale dependence of the integrated
      cross section for $p_T>30$\,GeV at $\mhiggs=120$\,GeV. (a)
      $\muR=\mhiggs$ fixed, $\muF$ varies --- (b) $\muF=\mhiggs$ fixed,
      $\muR$ varies --- (c) $\muF=\mhiggs/4$ fixed, $\muR$ varies ---
      (d) same as (b), but with cut on $|y|<2$.}
  \end{center}
\end{figure}

In \fig{fig::ptmin} we show the integrated cross section for $h+$jet
production with a minimum $p_T$ of the Higgs of $p_{T,\rm cut} =
30$\,GeV:
\begin{equation}
\begin{split}
\sigma(p_T>p_{T,\rm cut}) = \int_{p_T>p_{T,\rm cut}}\dd
  p_T\frac{\dd\sigma}{\dd p_T}
\end{split}
\end{equation}
at $\mhiggs=120$\,GeV as a function of (a) the factorization and (b,c)
the renormalization scale. The factorization scale dependence is already
quite small at \lo{} and improves slightly at \nlo{}. The
renormalization scale dependence we show for (b) $\muF=\mhiggs$ and (c)
$\muF=\mhiggs/4$, respectively. For both choices, the \nlo{} corrections
improve the scale dependence significantly, although $\muF=\mhiggs$
seems to lead to a more natural behaviour of the \lo{} and the \nlo{}
curves.  This behaviour does not change much if we restrict the Higgs
rapidity to $|y|<2$, as shown in \fig{fig::ptmin}\,(d).

\section{\nnlo{} cross section with \bld{p_T} cut}\label{sec::nnlo}

\begin{figure}
  \begin{center}
    \begin{tabular}{c}
      \subfigure[]{\includegraphics[width=.5\textwidth]{%
          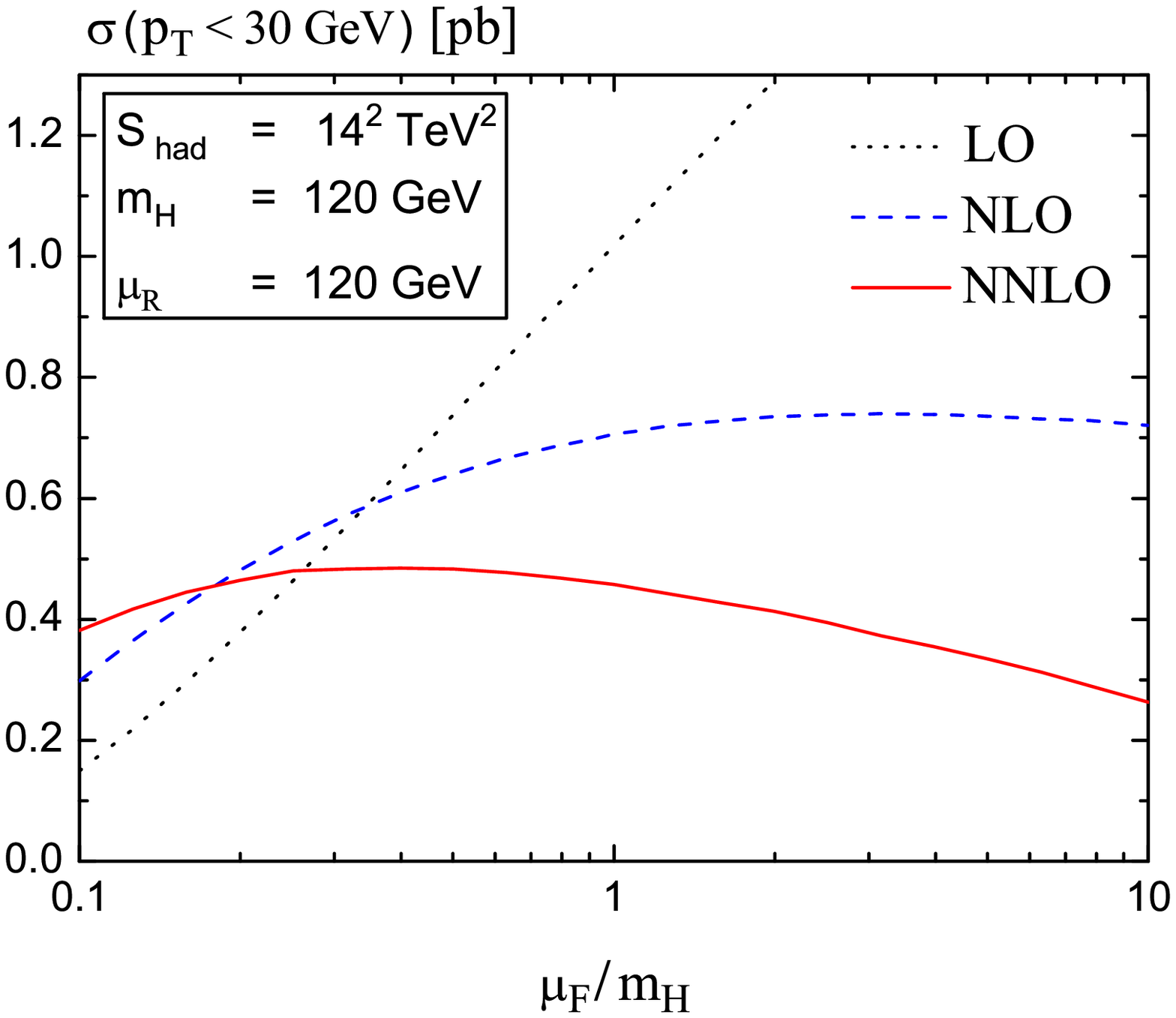}}
      \subfigure[]{\includegraphics[width=.5\textwidth]{%
          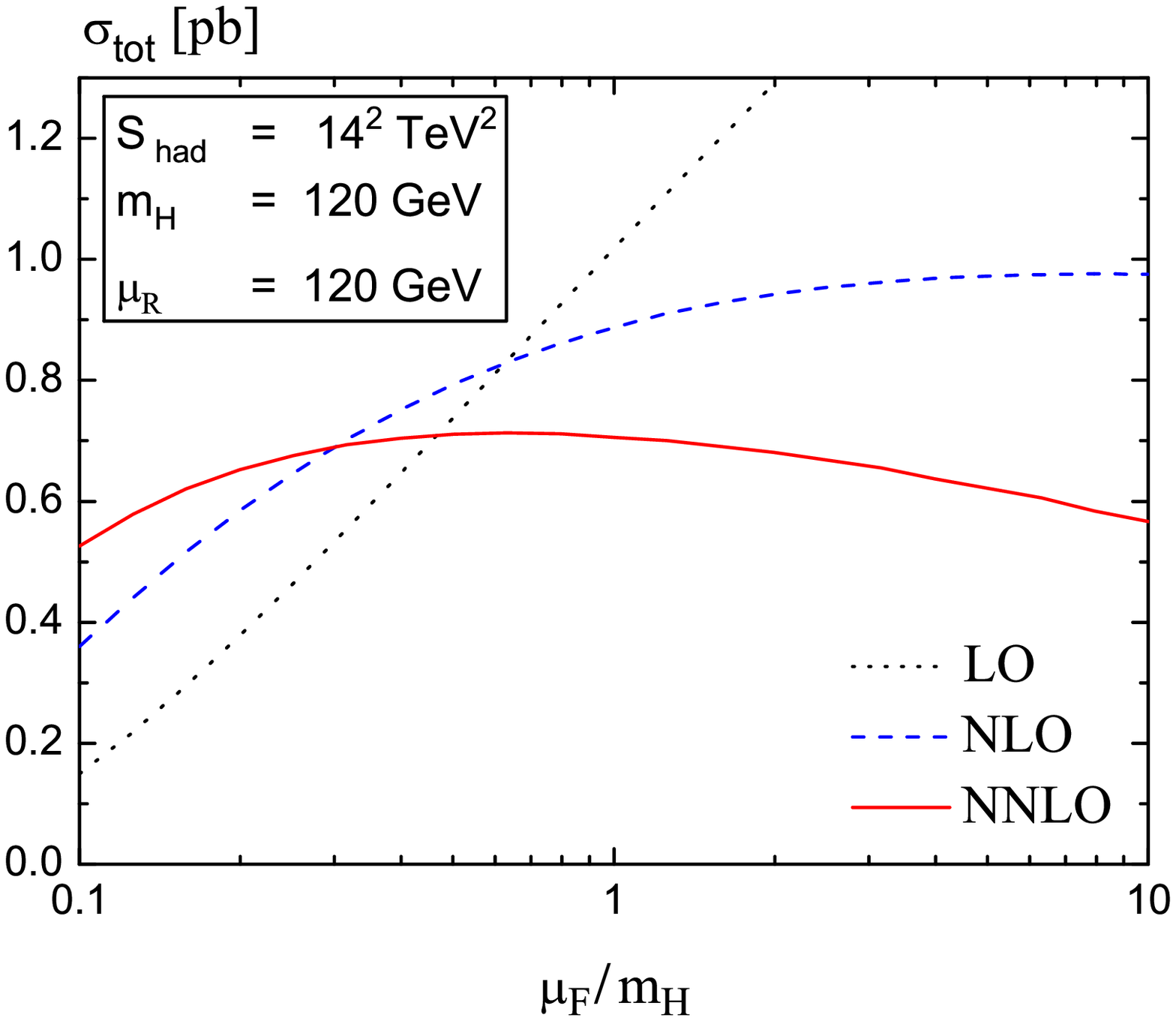}}\\
      \subfigure[]{\includegraphics[width=.5\textwidth]{%
          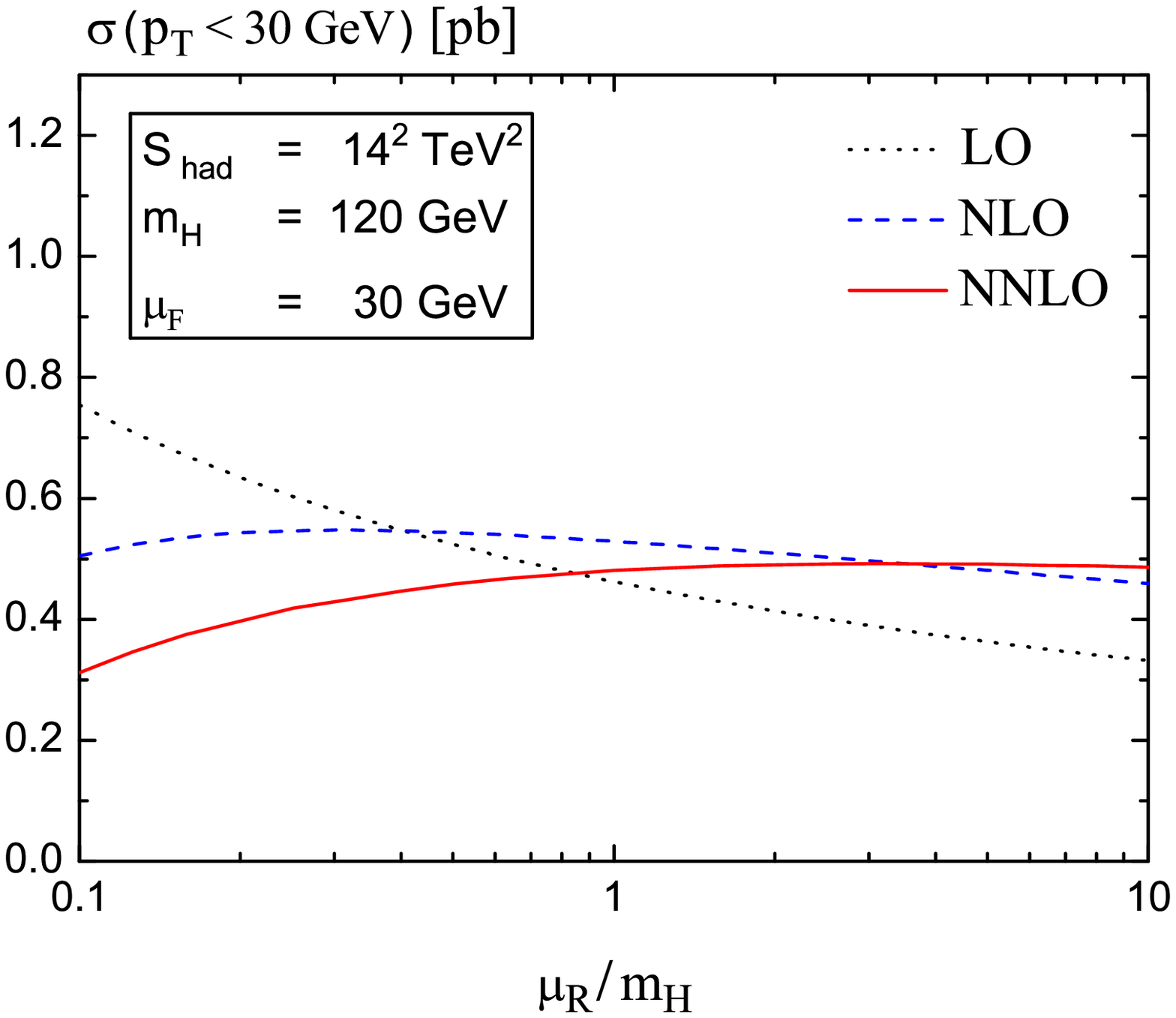}}
      \subfigure[]{\includegraphics[width=.5\textwidth]{%
          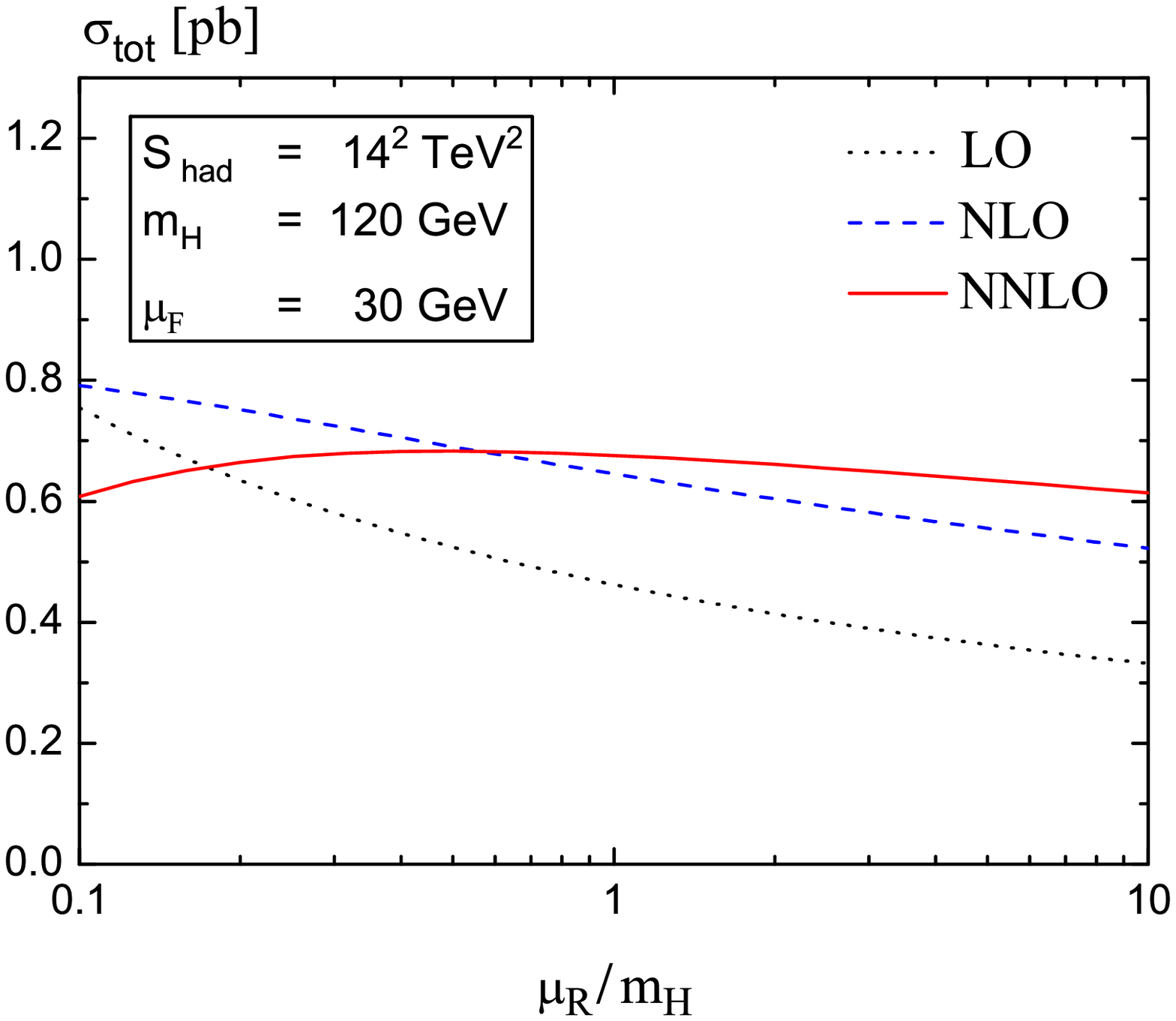}}
    \end{tabular}
    \caption[]{\label{fig::ptcut} Scale dependence of the integrated
      cross section for $p_T<30$\,GeV at $\mhiggs=120$\,GeV. (a)
      $\muR=\mhiggs$ fixed, $\muF$ varies --- (b) $\muF=\mhiggs/4$ fixed,
      $\muR$ varies.}
  \end{center}
\end{figure}

The knowledge of the total inclusive cross section $\sigma_{\rm tot}$ at
\nnlo{}~\cite{Harlander:2003ai} allows us to use the results of the
current paper to evaluate the inclusive cross section when applying a
finite $p_T$ cut:
\begin{equation}
\begin{split}
\sigma(p_T<p_{T,\rm cut}) = \int_{p_T<p_{T,\rm cut}}\dd
p_T\,\frac{\dd\sigma}{\dd p_T} = \sigma_{\rm tot} - \int_{p_T>p_{T,\rm
    cut}}\dd p_T\,\frac{\dd\sigma}{\dd p_T}\,.
\label{eq::ptcut}
\end{split}
\end{equation}
Of course, $p_{T,\rm cut}$ must not be too small in order not to be
sensitive to the region where large $\ln(p_T/\mhiggs)$ terms spoil
perturbative convergence. Furthermore, if $\sigma(p_T<p_{T,\rm cut})$ is
to be evaluated with \nnlo{} accuracy, we have to evaluate both terms
on the right side of \eqn{eq::ptcut} with \nnlo{} \pdf{}s.

\fig{fig::ptcut}\,(a) shows $\sigma(p_T<p_{T,\rm cut})$ for $p_{T,\rm
  cut}=30$\,GeV as a function of the factorization scale $\muF$, for
$\muR=\mhiggs$, varied over the rather large interval
$\muF\in[0.1,10]\mhiggs$.  The observations are very similar to the
fully inclusive cross section obtained without
cuts~\cite{Harlander:2003ai} which is shown for comparison in
\fig{fig::ptcut}\,(b): Perturbation theory prefers a scale significantly
below $\mhiggs$. With this choice, the renormalization scale dependence
is very weak around $\muR=\mhiggs$ already at \nlo{}, as shown in
\fig{fig::ptcut}\,(c) and \fig{fig::ptcut}\,(d).

\begin{figure}
  \begin{center}
    \begin{tabular}{c}
      \subfigure{\includegraphics[width=.7\textwidth]{%
          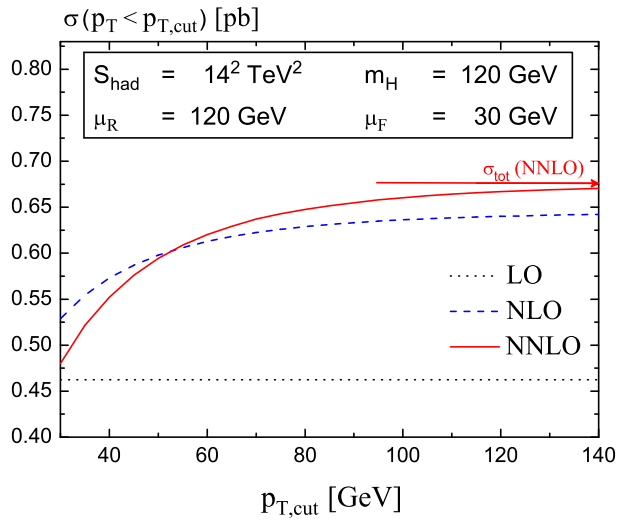}}
    \end{tabular}
    \caption[]{\label{fig::ptint} Inclusive cross section with an upper
      cut $p_{T,\rm cut}$ on the Higgs transverse momentum, see
      \eqn{eq::ptcut}, as a function of $p_{T,\rm cut}$. The dotted,
      dashed, and solid curves show the \lo{}, \nlo{}, and the \nnlo{}
      result. The arrow on the right indicates the value of the \nnlo{}
      result without cuts, $\sigma_{\rm tot}$.}
  \end{center}
\end{figure}

Finally, \fig{fig::ptint} shows the \nnlo{} cross section
$\sigma(p_T<p_{T,\rm cut})$ as a function of $p_{T,\rm cut}$. Since the
\lo{} only contributes at $p_T=0$, it is independent on $p_{T,\rm
  cut}$. The \nnlo{} corrections are negative with respect to \nlo{} at
small $p_{T,\rm cut}$ and change sign at around $p_{T,\rm cut}\approx
50$\,GeV.

\section{Conclusions and Outlook}

In this paper we have presented a first study of higher order
differential distributions for Higgs production in bottom quark
annihilation. The five-flavour scheme was used to calculate \nlo{} $p_T$
and $y$ distributions of the Higgs boson. Combination with the inclusive
\nnlo{} total cross section allowed us to derive the inclusive cross
section with upper cuts on the Higgs transverse momentum at \nnlo{}.

Concerning the choice of the factorization scale, we find that it
strongly depends on the observable under consideration: the value
$\muF=\mhiggs/4$~\cite{Plehn:2002vy,Maltoni:2003pn,Harlander:2003ai}
seems to be favoured in particular when the region $p_T=0$ is involved,
but is less motivated otherwise.

Although we expect that the results will be phenomenologically relevant
on their own, our approach should be useful also for an extension to a
fully differential \nnlo{} Monte Carlo program for the process $b\bar
b\to H+X$ along the lines of Ref.\,\cite{Catani:2007vq}.

\paragraph{Acknowledgments.}

We would like to thank M.~Kr\"amer for useful comments on the
manuscript. This work was supported by {\abbrev DFG} contract
HA~2990/3-1 and by {\abbrev BMBF} grant 05H09PXE.

\def\app#1#2#3{{\it Act.~Phys.~Pol.~}\jref{\bf B #1}{#2}{#3}}
\def\apa#1#2#3{{\it Act.~Phys.~Austr.~}\jref{\bf#1}{#2}{#3}}
\def\annphys#1#2#3{{\it Ann.~Phys.~}\jref{\bf #1}{#2}{#3}}
\def\cmp#1#2#3{{\it Comm.~Math.~Phys.~}\jref{\bf #1}{#2}{#3}}
\def\cpc#1#2#3{{\it Comp.~Phys.~Commun.~}\jref{\bf #1}{#2}{#3}}
\def\epjc#1#2#3{{\it Eur.\ Phys.\ J.\ }\jref{\bf C #1}{#2}{#3}}
\def\fortp#1#2#3{{\it Fortschr.~Phys.~}\jref{\bf#1}{#2}{#3}}
\def\ijmpc#1#2#3{{\it Int.~J.~Mod.~Phys.~}\jref{\bf C #1}{#2}{#3}}
\def\ijmpa#1#2#3{{\it Int.~J.~Mod.~Phys.~}\jref{\bf A #1}{#2}{#3}}
\def\jcp#1#2#3{{\it J.~Comp.~Phys.~}\jref{\bf #1}{#2}{#3}}
\def\jetp#1#2#3{{\it JETP~Lett.~}\jref{\bf #1}{#2}{#3}}
\def\jphysg#1#2#3{{\small\it J.~Phys.~G~}\jref{\bf #1}{#2}{#3}}
\def\jhep#1#2#3{{\small\it JHEP~}\jref{\bf #1}{#2}{#3}}
\def\mpl#1#2#3{{\it Mod.~Phys.~Lett.~}\jref{\bf A #1}{#2}{#3}}
\def\nima#1#2#3{{\it Nucl.~Inst.~Meth.~}\jref{\bf A #1}{#2}{#3}}
\def\npb#1#2#3{{\it Nucl.~Phys.~}\jref{\bf B #1}{#2}{#3}}
\def\nca#1#2#3{{\it Nuovo~Cim.~}\jref{\bf #1A}{#2}{#3}}
\def\plb#1#2#3{{\it Phys.~Lett.~}\jref{\bf B #1}{#2}{#3}}
\def\prc#1#2#3{{\it Phys.~Reports }\jref{\bf #1}{#2}{#3}}
\def\prd#1#2#3{{\it Phys.~Rev.~}\jref{\bf D #1}{#2}{#3}}
\def\pR#1#2#3{{\it Phys.~Rev.~}\jref{\bf #1}{#2}{#3}}
\def\prl#1#2#3{{\it Phys.~Rev.~Lett.~}\jref{\bf #1}{#2}{#3}}
\def\pr#1#2#3{{\it Phys.~Reports }\jref{\bf #1}{#2}{#3}}
\def\ptp#1#2#3{{\it Prog.~Theor.~Phys.~}\jref{\bf #1}{#2}{#3}}
\def\ppnp#1#2#3{{\it Prog.~Part.~Nucl.~Phys.~}\jref{\bf #1}{#2}{#3}}
\def\rmp#1#2#3{{\it Rev.~Mod.~Phys.~}\jref{\bf #1}{#2}{#3}}
\def\sovnp#1#2#3{{\it Sov.~J.~Nucl.~Phys.~}\jref{\bf #1}{#2}{#3}}
\def\sovus#1#2#3{{\it Sov.~Phys.~Usp.~}\jref{\bf #1}{#2}{#3}}
\def\tmf#1#2#3{{\it Teor.~Mat.~Fiz.~}\jref{\bf #1}{#2}{#3}}
\def\tmp#1#2#3{{\it Theor.~Math.~Phys.~}\jref{\bf #1}{#2}{#3}}
\def\yadfiz#1#2#3{{\it Yad.~Fiz.~}\jref{\bf #1}{#2}{#3}}
\def\zpc#1#2#3{{\it Z.~Phys.~}\jref{\bf C #1}{#2}{#3}}
\def\ibid#1#2#3{{ibid.~}\jref{\bf #1}{#2}{#3}}
\def\otherjournal#1#2#3#4{{\it #1}\jref{\bf #2}{#3}{#4}}
\def\jref#1#2#3{{\bf #1} (#2) #3}

\newcommand{\bibentry}[4]{#1, {\it #2}, #3.}
\newcommand{\hepph}[1]{{\tt hep-ph/#1}}
\newcommand{\hepth}[1]{{\tt hep-th/#1}}
\newcommand{\hepex}[1]{{\tt hep-ex/#1}}
\newcommand{\mathph}[1]{{\tt math-ph/#1}}
\newcommand{\arxiv}[2]{{\tt arXiv:#1}}

\end{document}